\documentclass[12pt]{amsart}
\usepackage{geometry}                
\usepackage{bm}
\usepackage{graphicx}
\usepackage{graphics}
\usepackage{float}
\usepackage{amsmath}
\usepackage{amssymb}
\usepackage{amsfonts}
\usepackage{latexsym}
\usepackage{epsfig}
\usepackage{dcolumn}
\usepackage{url}\usepackage{epstopdf}
\newcommand{\Ocal}{{\mathcal O}}

\newcommand{\Ecal}{{\mathcal E}}

\newcommand{\Ical}{{\mathcal I}}
\newcommand{\Hcal}{{\mathcal H}}

\newcommand{\Tcal}{{\mathcal T}}

\newcommand{\half}{{\textstyle \frac{1}{2}}}

\newcommand{\bvec}{{\textbf b}}

\newcommand{\dvec}{{\textbf d}}
\newcommand{\uvec}{{\textbf u}}

\newcommand{\rvec}{{\textbf r}}
\newcommand{\svec}{{\textbf s}}

\newcommand{\nvec}{{\textbf n}}

\newcommand{\jvec}{{\textbf j}}

\newcommand{\grad}{{\bm\nabla}}

\newcommand{\nuvec}{{\bm \nu}}
\newcommand{\sigmavec}{{\bm\sigma}}

\newcommand{\Avec}{{\textbf A}}
\newcommand{\Bvec}{{\textbf B}}

\newcommand{\Lvec}{{\textbf L}}

\newcommand{\rot}{{\nabla \times}}
\newcommand{\curl}{{\nabla \times}} 
\usepackage{epstopdf}

\usepackage{color}   

\usepackage[francais,english]{babel}
\usepackage{lmodern} 
\usepackage[T1]{fontenc}
\title{Tubes of Magnetic Flux and Electric Current \\
in Space Physics}
\author{Maurice Kleman\dag { }and Jonathan M. Robbins\ddag}

\begin{document}
\address{\dag Institut de Physique du Globe de Paris - 
 Sorbonne Paris Cité - 1 rue Jussieu, Paris cedex 05 France, \ddag Department of Mathematics, University of Bristol, University Walk, Clifton, Bristol BS8 1TW UK}
  \email{\textsf{{kleman@ipgp.fr}, j.robbins@bristol.ac.uk}}
\maketitle

\begin{abstract}
The singularities of an irrotational magnetic field are lines of electric current. This property derives from the relationship between vector fields and the topology of the underlying three-space and allows for a definition of {cosmic field} flux tubes and flux ropes as \textit{cores} (in the sense of the physics of defects) of helical singularities. {When applied to force-free flux ropes, and assuming current conservation,} an interesting feature is the quantization of the radii, pitches, and helicities. {One expects similar quantization effects in the general case.} In the special case when the total electric current vanishes, a force-free rope embedded in a medium devoid of magnetic field is nonetheless topologically stable, because it is the core of a singularity of the vector potential. 
Magnetic merging is also briefly discussed in the same framework.  
\end{abstract}

\section{Introduction}\label{s1}
    
    One of the still not fully understood features of  cosmic magnetic configurations is the universal presence of tubes of magnetic flux, \textit{i.e.}~filamentary domains in which the magnetic field is concentrated. Parker \cite{parker79} (chapter 10, p. 208) writes: {\color{black}``The concentration of field into isolated flux tubes occurs spontaneously, in opposition to the considerable magnetic pressure of the concentrated field. The phenomenon challenges our understanding of the basic physics of flux tubes.'' Parker goes on to discuss a number of physical mechanisms to which the stability of the flux tubes might be attributed. In a recent review  \cite{parker09}, written thirty years after the above reference,  
 his view is that the fibril state of solar magnetic fields remains unexplained.}
    
    Flux tubes were first observed in the Sun's corona on the occurrence of eclipses, and  have since been routinely investigated with coronagraphs that occult the solar disk and {reveal} the relatively low intensity photons emitted by the corona 
 and in soft X-rays.   
For a review of corona observations, see \cite{aschwanden06} (chapter~1). {Flux tubes have  also been observed  by crossing spacecraft as major structural features of the plasmoids in the Earth's magnetotail 
(\cite{schrijver11}, chapter 6).}
But there is some evidence that flux tubes are ubiquitous throughout the Universe on a variety of scales, in stars other than the Sun, in galaxies, and in intergalactic space. Thus, one expects that the flux tubes in the Sun provide a model for stars where ``starspots" have been observed \cite{parker09}. ``Twisted trunks" visible in some planetary nebulae have been interpreted as twisted filaments confined by a magnetic field 
{\cite{dahlgren07}}. Extragalactic filamentary magnetic networks in clusters of galaxies have also been discovered \cite{fabian08}.

    A number of computer simulations {have reproduced the major features of these tubes but without giving a clear understanding of their origin.}  It has been suggested that these flux concentrations are due to the expulsion of the magnetic field by eddies {\color{black} \cite{vishniac95, cattaneo03}}. This is certainly true, but is not the sole reason, we claim. This behavior has been compared to the expulsion of the magnetic flux in superconductors of the second species, see \cite{yoshida99}; according to their theory, {which assumes a completely ionized plasma}, {flux concentration occurs} only for a specific range of the {parameters} that enter the solutions of the MHD equations. However, the phenomenon is too universal to be limited to particular solutions. 
    
    We propose here a possible explanation that could work in all circumstances. The only prerequisite is that the cosmic magnetic field is  \textit{irrotational}  {\color{black} throughout most of the region of interest}, from which assumption we demonstrate that the singularities (the ``defects") of such a universal geometry are tubes of electric current  {\color{black}(similar to Birkeland currents {\bf\color{black}\cite{birkeland}})}, and consequently, thanks to the Biot$-$Savart relationship, tubes of high magnetic flux. Thereby, the tube stability is of a \textit{topological} nature, which is in no way connected to the presence of eddies. 
        
    {\color{black}The assumption of an irrotational magnetic field is frequently invoked in the heliophysics literature. Aschwanden \cite{aschwanden06} (chapter 5) stresses that the coronal magnetic field ``can be quantified in terms of a potential field which characterizes to first order unipolar fields in sunspots, dipolar fields in active regions, and is often used to compute the global coronal field with a source-surface model.''  At the same time it is recognized that there are regions of concentrated current, in the form of helically twisted loops, or tubes, 
  where the magnetic field reaches values in the thousand-gauss range, while outside these regions the field strength is typically less than 5\,\% of these values.  We shall see that it makes sense to assume that the magnetic field is in fact vanishing outside the tubes.}
    
    It is usual to distinguish between flux tubes with or without magnetic helicity $\Hcal =\int \Avec\cdot\Bvec\, \mathrm{d}V$. The tubes with non-vanishing $\Hcal$, the so-called \textit{twisted flux tubes} or {\textit{flux ropes}}, whose origin is still disputed (\textit{e.g.}, produced in the convection zone of the Sun, or resulting from photospheric flows, or $\dots$), are attended by coronal mass ejections {(CMEs)} and are {thought to be related to} the formation of prominences \cite{schrijver11}.  The question of the nature of a flux tube is equivalent to the question of the nature of the  \textit{core} of a singularity, \textit{i.e.} how the medium escapes the presence of a physically impossible singular solution that would otherwise be unavoidable on topological grounds. {\color{black} The core is a domain with properties that render the singularity virtual. Thus, to solve for a singularity is to solve a problem where two solutions are in contact at the core boundary. Such a question occurs frequently in the theory of defects in condensed matter physics, see \cite{friedeldisloc}.}
    
    {\color{black}The magnetic-field singularities that we discuss here are {distinct from} those arising from tangential discontinuities (current sheets), which are a necessary consequence of the equilibrium balance of Maxwell stresses \cite{parker94}, as well as the separatrices and null points that shape the global magnetic geometry at the surface of the Sun \cite{longcope05}}.
    
 In this article we shall stress the physical properties of flux ropes in relation to their topological properties. For the sake of simplicity, the calculations are developed for \textit{cylindrical} tubes and {\it force-free fields}, \textit{i.e.}~the magnetic stresses are in equilibrium with themselves: {\color{black}Sections~\ref{s4} and \ref{trnf}}. We also assume that the current is conserved; this is not the most general assumption for cosmic tubes, but it has been observed in some situations occurring in turbulent plasmas where the total current is indeed a determining parameter \cite{suzuki03}. However we discuss in Section~\ref{s8} some other parameters (flux, helicity) when they are determining, and an article is in preparation devoted to the fixed-flux case. These simplifications show the importance of the consideration of topological features in more general cases.
Some considerations applying to the general case are given in Section~\ref{singtheory}.

  \section{Tubes of Electric Current are Topological Defects}\label{s2}
  Let us assume the existence of a magnetic field $\Bvec$ with no associated current ($\rot\Bvec = 0$) deriving from the following scalar potential:
  \begin{equation}
\label{e1}
\psi \equiv B_0\left(z +\frac{b}{2\pi}\theta\right).
\end{equation}The isopotential surfaces are ruled helicoids and the magnetic field,  expressed in cylindrical coordinates
{$(r,\theta,z)$}, is given by
  \begin{equation}
\label{e2}
\Bvec =\grad \psi = \{0, B_0\frac{b}{2\pi r}, B_0\}.
\end{equation} 
{\color{black} We wish to investigate}   the physical nature of the singularity of the field along the $\mathbf{\hat z}$-axis.

Here and in what follows, 
certain developments are inspired by analogies with the theory of defects in condensed-matter physics.  While our results do not depend explicitly on results from this theory, we shall comment briefly on some of these analogies, as they may be useful and suggestive. For example,
a helical geometry similar to that of Equation (\ref{e2})  appears in \textit{screw dislocations} in liquid crystalline \textit{smectic phases}. In the ground state,  these phases consist of planar equidistant lamellae; they can suffer certain singularities \cite{kln} such as these 
screw dislocations, which are among the simplest ones, 
in which the lamellae are helicoids
(for completeness,  we include a brief discussion of defects in condensed matter  in the appendix).
Let $d$ be the {\color{black} distance} between lamellae in the ground state of a smectic phase. {It follows that the pitch $p$ of the helicoids, {which are given by $\psi_{\mathrm {lam}} = z \mp (nd/2\pi)\theta = 0, \ n \in \mathbb{Z}$}, is a multiple $p=\pm nd$ of this distance.  The pitch can be expressed} as an integral along a loop $C$ (the \textit{Burgers' circuit}) surrounding the $\mathbf{\hat z}$-axis ({the locus of the dislocation}), namely
$$\oint_C \uvec\cdot\ \mathrm{d}\svec=  \Delta\psi_{\mathrm {lam}} =p,$$ where $\uvec = \grad \psi_{\mathrm {lam}}  = \{0,\mp nd/2\pi{r}),1\}$; {$p \,\bf\hat z$ is then the \textit{Burgers' vector}.} 

Analogously, consider the integral $\oint_C \Bvec\cdot \mathrm{d}\svec$ along a circle $C$ of radius $r$. One gets
\begin{equation}
\label{e3}
\oint_C \Bvec\cdot \mathrm{d}\svec = B_0 b,
\end{equation} which is independent of the circuit $C$ surrounding the $\mathbf{\hat z}$-axis.  This quantity measures the current intensity $\Ical$ through $C$, namely $(4\pi/c)\Ical = \iint \rot\Bvec\cdot \mathrm{d}\sigmavec$; $\rot\Bvec$ vanishes almost everywhere, so {\color{black} that} 
\begin{equation}
\label{e4}
\color{black} \rot\Bvec =\frac{4\pi}{c} \Ical \, \delta (x)\, \delta(y) \ \bf\hat{z},
\end{equation}
and, by employing Stokes' theorem,
\begin{equation}
\label{e5}
\frac{4\pi}{c} \Ical = B_0 b. \end{equation} {\color{black} The current intensity is concentrated along the $\mathbf{\hat z}$-axis (and is the equivalent of the Burgers' vector). On the boundary of the core region we assume }
$(\Bvec^+ -\Bvec^- )\cdot \nvec=0$ and $ \Avec^+-\Avec^-=0$ (see \cite{griffiths99}, chapter 5) and the global condition 
 \begin{equation}
\label{e6}
2\pi\int_0^{r_0}r\mathrm{d}r \ \mathbf{\hat z} \cdot\rot \Bvec = \frac{4\pi}{c} \Ical =B_0b.
\end{equation} Other conditions may apply in the resolution of the fundamental equations of magnetohydrodynamics, but we shall restrict  here {to steady-state solutions with no material current, \textit{i.e.}~to a purely magnetostatic problem.} We also assume that the medium is a perfect conductor. Finally, we impose cylindrical symmetry, so that $ \Bvec$ depends only on $r$. \\ \vspace{-10pt}
 
 We make the following remarks:
 
 i) The current along the $\mathbf{\hat z}$-axis is of course the source of a magnetic field [$\Bvec$] through the Biot$-$Savart relation. (More precisely, the azimuthal component of the magnetic field is generated by the $z$-component of the current, and the azimuthal component of the current contributes to $B_z$ inside the flux tube.  But the (constant) $z$-component of $\Bvec$ outside the flux tube is not generated by the current, but is generated by external currents that are far away. We shall later consider a situation with no magnetic field outside the tube of current, in which case the Biot$-$Savart relation completely determines the field inside the tube.)
This is not the way we have introduced it; we have instead introduced a set of surfaces orthogonal to an irrotational magnetic field. Thus the current appears as a singularity of this field, making the current secondary to the field; this is in agreement with Parker's conception that it is the magnetic field that causes the currents, not the other way around, see \cite{parker07}.
At this point, we wish to stress that such a set of surfaces cannot be defined unless the magnetic field {has zero curl}; {a sufficient condition for the existence of a  set of surfaces orthogonal to a given magnetic field $\Bvec$
 is precisely that $\Bvec$ be irrotational;} in effect, $\rot\Bvec = 0 \Rightarrow \nuvec\cdot\rot\nuvec=0$, where $\Bvec = |\Bvec|\nuvec$. 

ii) The preceding results are \textit{topological} in essence; {\color{black} this is an important feature from the theory of defects in condensed-matter physics. Thereby the} constancy of $nd$ in the lamellar case and of $\Ical$ in the electric-current case are physical properties that have to be maintained whatever the deformation of the helicoidal surfaces with respect to the regular helicoids may be, provided the helicoidal topology is preserved. In the electric-current case, $\Ical$ being given, one can smoothly bring the actual potential surfaces, in the vicinity of a straight line current, to the shape of regular helicoids (which ensures $b= \mathrm{constant}$ and $B_0=\mathrm{constant}$), such that Equation (\ref{e5}) is satisfied.

iii) 
{The analogy between the topological defects in smectics and the topological defects of an irrotational magnetic field is restricted to screw dislocations, {but merits discussion because it yields the main characteristics of line defects in magnetic fields, in particular their helical features}. Furthermore, magnetic configurations in the cosmos {can exhibit} current sheets \cite{parker94}, which have no analogs in smectics.
 Magnetic-defect configurations in ferromagnetic materials, with domain walls and Bloch and N\'eel lines, are somehow topologically analogous, but with nothing comparable to the presence of electric currents \cite{klemag}. On the other hand, vortex lines (vortex sheets) in irrotational fluids are topologically analogous to current intensity lines (current sheets).}

\section{The Electric Current as a Singularity of the Magnetic Field}\label{singtheory}

The analysis that follows is an application of the relationship between the topology of a three-dimensional domain $\Omega$ and the calculus of vector fields defined on such a domain -- {\color{black} see \cite{cantarella02} for a physics-oriented review.  Let $\Bvec$ be a (divergenceless) magnetic field on $\Omega$ satisfying  {\it tangent  boundary conditions}, \textit{i.e.} $\Bvec_n = 0$, where $\Bvec_n$ denotes the normal component of $\Bvec$ on the boundary $\partial \Omega$.  If $\Bvec$ is curl-free, it follows that $\Delta \Bvec = 0$, where $\Delta$ is the Laplacian.  In this case, $\Bvec$ is called a {\it harmonic knot} \cite{cantarella02}.}

{\color{black} It turns out that harmonic knots are closely related to the topology of $\Omega$.}  More precisely, the space of such fields is finite dimensional, with dimension given by the genus $g$ of the boundary $\partial \Omega$ (or, if $\partial \Omega$ consists of a number of components, the sum of the genera of its components).  Indeed, harmonic vector fields can be characterised by a set of  $g$ invariants, either fluxes or line integrals, as follows: 
Construct a minimal set of surfaces $\Sigma_j$ in $\Omega$ such that $\Omega-\cup \Sigma_j$ is simply connected 
--  in the language of defects, these are ``cut surfaces"; mathematically, they are Seifert surfaces.
The number of cut surfaces is given by  $g$. The intersections of the $\Sigma_j$s with $\partial \Omega$ yield $g$ closed curves, $\partial \Sigma_j$, on the boundary $\partial \Omega$. One can also define $g$ additional closed curves $C_k$ on $\partial \Omega$ such that $C_k$ intersects $\partial \Sigma_k$ once and does not intersect any of the other $\partial \Sigma_j$s. Together, the $\partial \Sigma_j$s and $C_k$s constitute a basis for the first homology group of the boundary $\partial \Omega$ which therefore has dimension $2g$. 

Then, and this is the basic point which allows one to make contact with the theory of defects, a harmonic magnetic field $\Bvec$ is 
uniquely determined by its line integrals around the $C_k$s (which may be interpreted in terms of currents outside $\Omega)$, or, equivalently, by its fluxes through the cut surfaces $\Sigma_j$.  As explained below, there are physical reasons to emphasise the first determination, namely
\begin{equation} \label{e13}
   \beta_k= \oint_{C_k} \Bvec\cdot \mathrm{d}\svec.
\end{equation}

In simple examples this construction is easily visualised.  For example, \cite{mahajan98}, for the purpose of modeling a tokamak, take $\Omega$ to be a solid torus $\Tcal$; $\Sigma$, a cross-section of the torus; $\partial \Sigma$, a closed curve going round the torus the ``short way''; and $C$, a closed curve going around the torus the ``long way''. In this case, $g = 1$.

The main example that we consider, also with $g = 1$, is directly related to the notion of defect developed in this article, and is shown in Figure~1.  The domain 
$\Omega$ is obtained by removing a solid torus $\Tcal$ from the interior of a (deformed) three-dimensional ball.   The boundary $\partial \Omega$ consists of a simply connected outer component $\partial \Omega_{\mathrm{out}}$ which is just the boundary of the three-dimensional ball, and an inner component $\partial\Omega_{\mathrm{in}} = \partial \Tcal$.  In order to render $\Omega$ connected, we introduce a cut surface $\Sigma$ whose boundary $\partial \Sigma$ is  a closed curve that goes around $\Tcal$ the ``long way''.  For ease of notation we will denote $\partial \Sigma$ by $L$.  We also introduce a closed curve $C$ which goes around $\Tcal$ the ``short way'', which intersects $L$ once.
  \begin{figure}[h]
  \centering
  \includegraphics[width=3.5in]{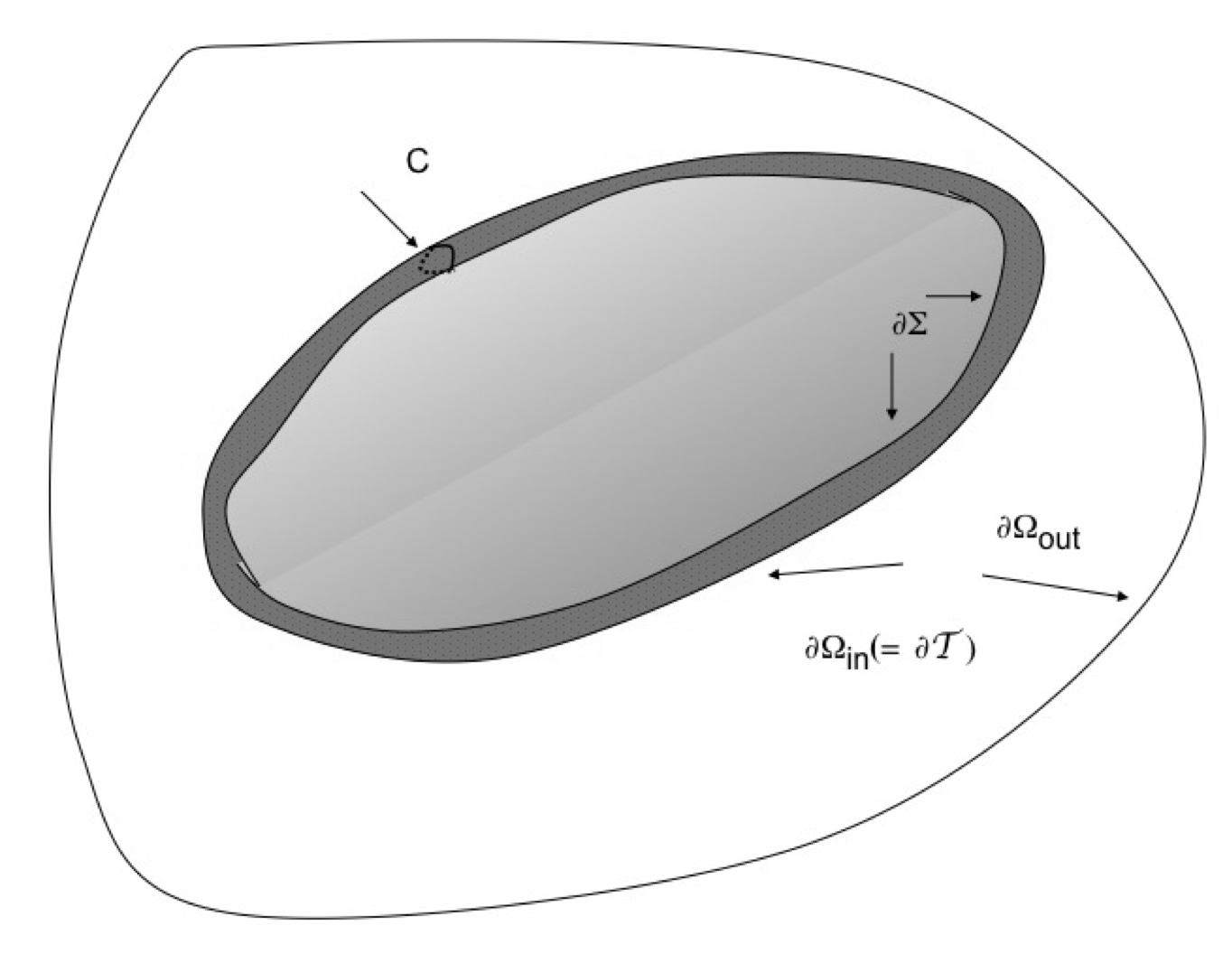}
\caption{\small A domain $\Omega$ with $g=1$. The colored torus is supposed to be empty (this restriction can be lifted; then the interior of the torus is the core of the singularity). $\partial \Sigma$ is along the ``long way'', and $C$ along the ``short way''. The cut surface $\Sigma$ is shadowed.}
\label{fig1}
\end{figure} 
\indent The quantity $\beta$ given by  Equation (\ref{e13}) takes the same value if $C$ is replaced by a homologous loop $C'$ since, by Stokes theorem,
the difference $\oint_C \Bvec \cdot \mathrm{d}\svec -\oint_{C'}\Bvec \cdot \mathrm{d}\svec$ is given by the integral of $\grad \times \Bvec$ over a surface in $\Omega$, and $\grad\times\Bvec$ vanishes by assumption. {\color{black} This invariance gives 
 $\partial\Tcal$ the status of a line defect, inasmuch as the radius of $\Tcal$ can be reduced to zero in such a way that $\Tcal$ collapses onto the closed curve $L$.  In this limit,  $\Bvec$ becomes singular on $L$. (In the terminology of the appendix,  $L$ corresponds to a vortex line, with $\Sigma$ as cut surface, and $C$ corresponds to a Burgers' circuit).} 

We may regard $\beta$ as proportional to a current $\Ical$ threading the interior of the torus $\Tcal$. Indeed, in the limit that $\Tcal$ collapses to $L$, we may write that
\begin{equation}
\label{e14}
\rot \Bvec =\frac{4\pi}{c}\ \Ical \hat \Lvec\, \delta^2_{ L}(\rvec),
\end{equation}
where $\hat\Lvec$ is unit-vector along $L$ and $ \delta^2_{ L}(\rvec)$ denotes the normalised two-dimensional $\delta$-function on $L$.

As $\Bvec$ is irrotational, it may be expressed (locally) as the gradient of a (multi-valued) scalar field $\psi$.  Introducing local cylindrical coordinates $(r,\theta,z)$ around $L$, we may  write
\begin{equation}
\label{e15}
\psi = \frac{\beta}{2\pi}\theta +\gamma z + \Ocal(r^2)+\mathrm{higher\ order\ terms}.
\end{equation}
There is no linear term in $r$, since $\Bvec_n =0$ on $\partial \Tcal$.   Equation (\ref{e1}) is a particular instance of  Equation (\ref{e15}), where $\Omega$ is the solid cylinder with its axis removed (and the higher-order terms in Equation (\ref{e15}) are absent).  

Finally, we note that provided the current $\Ical$ remains fixed, $\beta$ is invariant under deformations of $L$.  This is in contrast to the alternative invariant, namely the flux of $\Bvec$ through the cut surface $\Sigma$, which depends not only on $\Ical$ but also on the geometry of  $\Omega$.  This definition is adapted to a situation where the current-tubes are mobile while keeping their current constant, in which case $\beta$ appears as the characteristic invariant of a harmonic magnetic field $\Bvec$.

In the following sections, where we restrict to straight tubes of flux, the full characteristics of this general theory of singularities in an irrotational magnetic field will not be required, and it will be sufficient to think of the tubes as the cores of helical defects with regular helical level sets $\psi$. 

 \section{The Twisted Flux Rope}\label{s4}

Static configurations of currents and fields may be obtained by requiring the Lorentz force to vanish, \textit{i.e.} 
 $\jvec\times\Bvec = 0$. Hence we consider cores with force-free configurations.
 This is a particular model of a \textit{flux rope} in the sense of Parker (\cite{parker79}, chapter 9), although Parker assumes that there is no magnetic field outside the tube; we consider this case, which yields the most physically sensible results, in Section~\ref{trnf}. The force-free field situation is plausible for the magnetic ropes of the corona, where the {\color{black} plasma-$\beta$ --  that is, the ratio of the plasma gas pressure, $n_e k_B T$, to the magnetic pressure, $B^2/8\pi$ -- is very small,} so that the pressure gradient is negligible compared to the Lorentz force, which therefore must vanish in a magnetostatic situation. {The condition for a force-free field is given by the Beltrami equation:   
\begin{equation}
\label{e16}
 \ell \rot\Bvec =  \Bvec.
\end{equation}
{\color{black} We consider for the sake of simplicity the case where  $\ell$ is constant.  The form of the exterior field, Equation~(\ref{e2}),  dictates that core field  is axisymmetric.  
 The axisymmetric solution of Equation~(\ref{e16}) in a cylindrical domain} is well known since Lundquist  (see \cite{alfven63a}); it can be written in terms of Bessel functions of the first kind:
 \begin{equation}
\label{e17}
B_\theta = A J_1\left(\frac{r}{\ell}\right), \qquad B_z = A J_0\left(\frac{r}{\ell}\right).
\end{equation}
Requiring $\Bvec$ to be continuous on the boundary $r = r_0$, we obtain
 \begin{equation}
\label{e18}
 A  J_1(\eta) = \frac{B_0 b}{2\pi r_0}, \qquad A J_0(\eta) = B_0,\end{equation} 
where $\eta = r_0/\ell$, which implies that
 \begin{equation}
\label{e19}
\frac{J_1(\eta)}{J_0(\eta)}=\frac{b}{2\pi r_0}.
\end{equation}}

We assume that the two essential invariants of the flux rope, namely the current intensity $\Ical =(c/4\pi)B_0 b$ and the magnetic field $B_0$, are given; $\ell$ appears as a function of $r_0$ through the boundary condition Equation (\ref{e19}), and is an unknown of the problem, which is to be determined, as is $r_0$, by minimizing the energy.  {In what follows, without any essential loss of generality, we may restrict attention to  $\ell>0$, while $b$ may be of either sign.}

\begin{figure}[h] 
   \centering
   \includegraphics[width=3.5in]{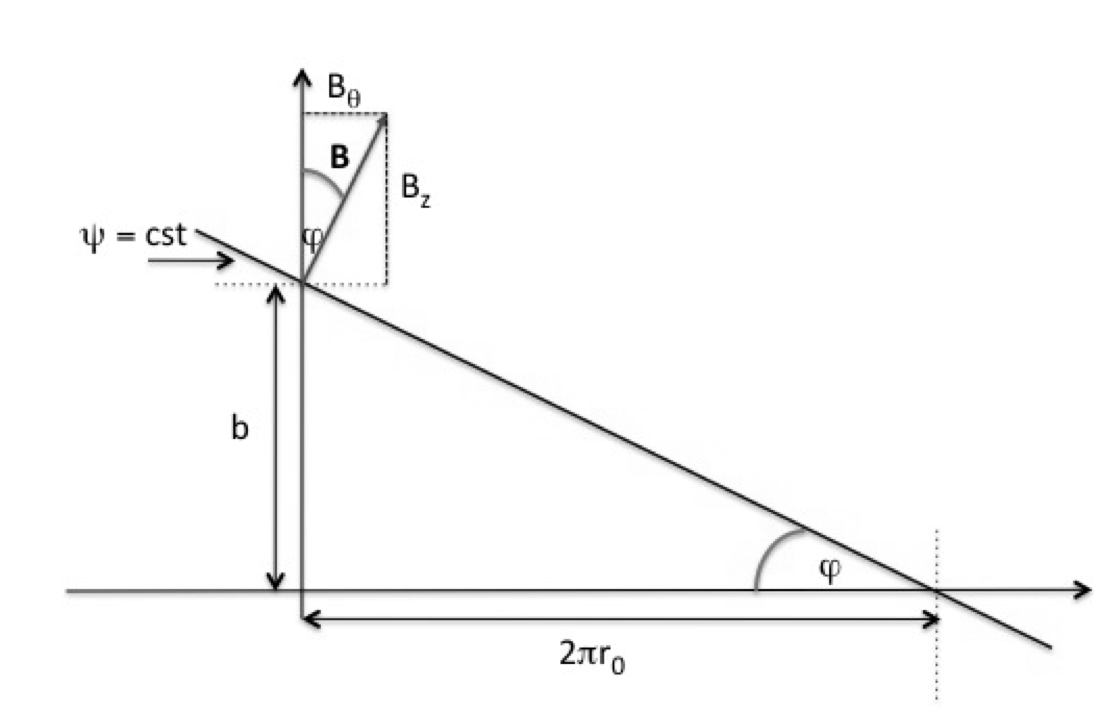} 
   \caption{\footnotesize The helical intersection of an helicoid $\psi =\mathrm{constant}$ with the boundary of the tube, $b>0$. Development on the plane.}
   \label{fig2}
\end{figure}

\begin{figure}[h] 
    \centering
    \includegraphics[width=4.in]{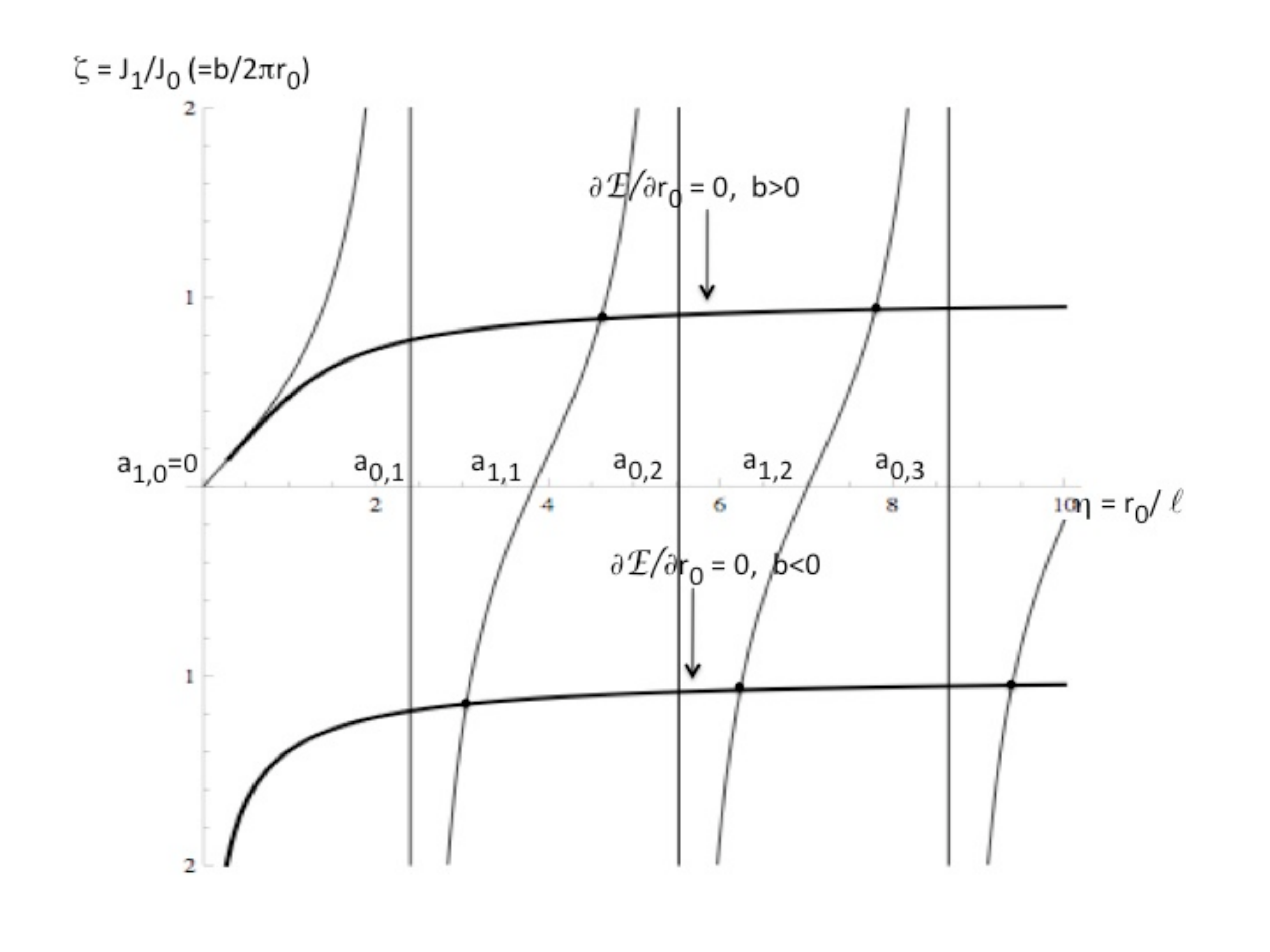} 
    \caption{\footnotesize  Nonvanishing magnetic field outside the rope and continuous at the boundary, with $\ell > 0$.  {\color{black}Thin} curves: boundary conditions at $r=r_0$; {\color{black} thick} curves: plot of the minimization condition $\partial \Ecal /\partial{r_0} = 0$, $\zeta>0 \ \mathrm{for}\  b>0$, $\zeta<0 \ \mathrm{for} \ b<0$. See Tables~\ref{table1}  and \ref{table2} for numerical values.}
    \label{fig3}
 \end{figure}

Notice that, when navigating around the cylinder one turn, \textit{i.e.} a distance $2\pi r_0$, a representative point at $z=z_0$ is lifted to ${\color{black}z_0}+ b$ along a geodesic of the cylinder; see Figure~\ref{fig2}. Thus $\tan\phi = J_1(\eta)/J_0(\eta)$, where $\tan\phi$ is the slope of this geodesic, which develops into a helix with pitch $-b$ on the cylinder $r=r_0$. Therefore the lines of force of the magnetic field, which are orthogonal to this helix, are helices of {\textit{opposite sign}. If \textit{e.g.}~the helicoids are left-handed ($b>0$), the magnetic field follows right-handed helices
of pitch $p(r_0) = (2\pi r_0)^2/b$ at $r=r_0$. For any $r<r_0$, the pitch of the helical lines of force of the magnetic field is given by $p(r) = 2\pi r J_0(r/\ell)/J_1(r/\ell)$.} {Thus, depending on the value of {\color{black} $r/\ell$}, the pitch may alternate in sign inside the cylinder.  Indeed, if we let $a_{0,\jmath}$, $j = 1,2,3,\ldots$, denote the zeros of $J_0$,  and $a_{1,\jmath}$ similarly denote the zeros of $J_1$, then $p(r)$ is positive on the intervals $\mathrm{a_{0,\jmath}} < r/\ell < \mathrm{a_{1,\jmath}}$ and negative on the intervals $\mathrm{a_{0,\jmath+1}} < r/\ell <  \mathrm{a_{1,\jmath+1}}$ (note that the zeros of $J_0$ and $J_1$ are interlaced, \textit{i.e.}~$a_{0,\jmath} < a_{1,\jmath} < a_{0,\jmath + 1}$, and that the magnetic field is horizontal at $r/\ell = a_{0,\jmath}$ and vertical at $r/\ell = a_{1,\jmath}$).
See Figures~\ref{fig3}, \ref{fig4}, and \ref{fig5} hereunder.}

The number of sign changes of the pitch is even or odd according to the relative sign of the interior and exterior fluxes. 
 Indeed, outside the tube, the magnetic flux per unit area is given by $B_0$. The flux inside the core $\Phi_i = \iint_0^{r_0} \Bvec(\rvec)r\,\mathrm{d}r\,\mathrm{d}\theta$ follows from the force-free condition and Stokes' theorem
\begin{equation}
\label{e20}
\Phi_i = {b\ell}B_0.
\end{equation} 
The interior and exterior fluxes have the same sign if $b>0$ and opposite signs if $b<0$. 
At the centre of the core, the pitch of 
the magnetic lines of force is necessarily positive (since $\ell > 0$).  For $b > 0$, 
the pitch of the magnetic lines of force is positive at the boundary; therefore, if the pitch changes sign in the core, it does so an even number of times.    For $b<0$, the pitch of the magnetic lines of force is negative at the boundary; it therefore necessarily changes sign inside the core (an odd number of times). 

We remark that $\ell$ can be given a simple interpretation {\color{black} (see \textit{e.g.} \cite{sakurai79})}. Because 
$$\frac{1}{\ell}=\frac{1}{|\Bvec|^2}\Bvec\cdot\curl\Bvec =\nuvec\cdot\curl\nuvec,\quad \mathrm{where} \ \ \Bvec = \nuvec | \Bvec |,$$ $\varpi=-2\pi\ell$ is the pitch of the vector field [$\nuvec$], which thus rotates helically with a constant velocity along the radii of the cylindrical geometry. One can say that $\Bvec$ suffers a \textit{double-twist}, one along $\bf{\hat z}$, the other along $\rvec$,  {\color{black} which is analogous to the double-twist observed in liquid crystalline blue phases and cholesteric phases, where in general $\varpi$ depends on $r$.} Of course $\varpi$,  {\color{black}hence $\ell$}, can also depend on $r$ in the MHD case, and {in a more general geometry} probably does, as a rule, without breaking the force-free condition.


The magnetostatic energy {\color{black} of the twisted flux rope} is the sum of the exterior and core contributions, 
$$\Ecal_{out}= \frac{1}{8\pi} \int_{r_0}^R r\, \mathrm{d}r \int_0^{2\pi}\mathrm{d}\theta\, (B_z^2 + B_\theta^2), \qquad \Ecal_{in}= \frac{1}{8\pi} \int_{0}^{r_0} r\, \mathrm dr \int_0^{2\pi}\mathrm d\theta\, (B_z^2 + B_\theta^2)$$ \textit{i.e.} \begin{equation}
\label{e21}
 \Ecal_{out} =\frac{B_0^2}{4}\left(\half (R^2-r_0^2) + \left(\frac{b}{2\pi}\right)^2\ln \frac{R}{r_0}    \right)
\end{equation}
and 
\begin{equation}
\label{e22}
\Ecal_{in} = 
\frac{B_0^2}{4}\left(r_0^2 +\left(\frac{b}{2\pi}\right)^2 -\frac{b\ell}{2\pi} \right),
\end{equation} 
where we have used the identities \cite{abram}

\begin{equation}
\label{e23}
  \int_0^\eta \rho \, \mathrm{d}\rho \, J_0^2(\rho)   = \frac12 \eta^2\left(J_0^2(\eta) + J_1^2(\eta)\right),\nonumber   \end{equation}
   \begin{equation} \label{e24}
   \int_0^\eta \rho \, \mathrm{d}\rho  \, J_1^2(\rho)   = \frac12 \eta^2\left(J_0^2(\eta) + J_1^2(\eta) - 2\frac{J_0(\eta)J_1(\eta)}{\eta}\right). 
\end{equation}
We obtain
\begin{equation} \label{e25}\Ecal= \frac{B_0^2}{8}\left(R^2 + r_0^2 + \frac{b^2}{2\pi^2}\ln \frac{R}{r_0} + \frac{b^2}{2\pi^2} - \frac{b\ell}{\pi}\right).\end{equation}
 
{We assume that $b$ and $B_0$ are independent invariants, and seek to minimize the total energy $\Ecal$ with respect to $r_0$. Recall from Equation (\ref{e19}) that the Beltrami length $\ell$ depends on $r_0$; for convenience, we write the relation in the form
\begin{equation}
\label{e26} F \equiv r_0 \ \frac{J_1}{J_0}(r_0/\ell) - \frac{b}{2\pi } =0.
\end{equation}
By the chain rule, one has
\begin{equation} \label{e27} {\partial\ell}/{\partial r_0}=-({{\partial F}/{\partial r_0}})/({{\partial F}/{\partial \ell}}). \end{equation}
Noting that
\begin{equation}\label{e28}
\frac{\partial F}{\partial \ell}(r_0) =-\frac{r_0^2}{\ell^2} 
\left(\frac{J_1}{J_0}\right)'(r_0/\ell),\quad
  \frac{\partial F}{\partial r_0}(r_0) =\left(\frac{J_1}{J_0}\right)(r_0/\ell) +\frac{r_0}{\ell} 
  \left(\frac{J_1}{J_0}\right)'(r_0/\ell) ,\end{equation}
 and that
 \begin{equation} \label{e29} 
 \left(\frac{J_1}{J_0}\right)' (r_0/\ell) =1 -\frac{\ell}{r_0}\left(\frac{J_1}{J_0}\right)(r_0/\ell) +\left(\frac{J_1}{J_0}\right)^2(r_0/\ell),
 \end{equation}  one obtains, after replacing ${J_1}/{J_0}$ by  (\textit{cf.}~Equation (\ref{e26})) 
 \begin{equation}\label{e30}
 \zeta = \frac{b}{2\pi r_0}
 \end{equation}
 the following relation:
  \begin{equation}
\label{e31}
\frac{\partial\ell}{\partial r_0}=-\frac{\zeta+\zeta^{-1}}{1-(r_0/\ell)(\zeta+\zeta^{-1})}.
\end{equation}}

The extrema of $\Ecal$ are given by $\partial \Ecal /\partial{r_0} = 0,$ \textit{i.e.}
\begin{equation}
\label{e32}
\qquad \frac{\partial \Ecal }{\partial{r_0}} = \frac{ B_0^2}{4}\, r_0\left(1+\zeta\frac{\zeta+\zeta^{-1}}{1-(r_0/\ell)(\zeta+\zeta^{-1})} -\zeta^2\right)= 0.
\end{equation}
After some easy algebraic manipulation, this equation can be written:
\begin{equation}
\label{e33}
{\frac{b}{2\pi\ell} = \frac{2\zeta^2}{1-\zeta^4}}.
\end{equation}
{It follows that if $b > 0$, then $0 < \zeta < 1$, while if $b < 0$, then $\zeta < -1$.}
For $\zeta=\pm 1$ one finds that $\ell = 0$: there is no twist.

The relation between $r_0$ and $\ell$ can be better understood by  consideration of Figure~\ref{fig3}. {The {\color{black} thin} curves are plots of 
the boundary condition $\zeta = J_1(\eta)/J_0(\eta)$ (\textit{cf.}~Equation (\ref{e26})), and are asymptotic to the vertical lines $\eta= \mathrm{a_{0,\jmath}}$. {\color{black}The thick curves} are plots of the minimisation condition Equation (\ref{e33}), expressed in the equivalent form
\begin{equation}
\label{e34}
\eta=\frac{2\zeta}{1-\zeta^4}.
\end{equation}
The upper {\color{black} thick curve} corresponds to $b>0$.  It is tangent to the graph of $J_1(\eta)/J_0(\eta)$ at $\eta = 0$, and has asymptote 
$\zeta = 1$.   Thus, for $\eta$ large, one gets solutions
$r_0 \approx {b}/{2\pi}$, whatever the value of $\ell$ may be (see Table~\ref{table1}). The lower {\color{black} thick curve} has asymptote  $\zeta=-1$.  Thus, for $\eta$ large, one has solutions $r_0 \approx -{b}/{2\pi}$  (see Table~\ref{table2}). The case $\ell<0$ is illustrated Figure~\ref{fig4}; the symmetries with respect to Figure~\ref{fig3} are obvious.
\begin{figure}[h] 
    \centering
    \includegraphics[width=4.in]{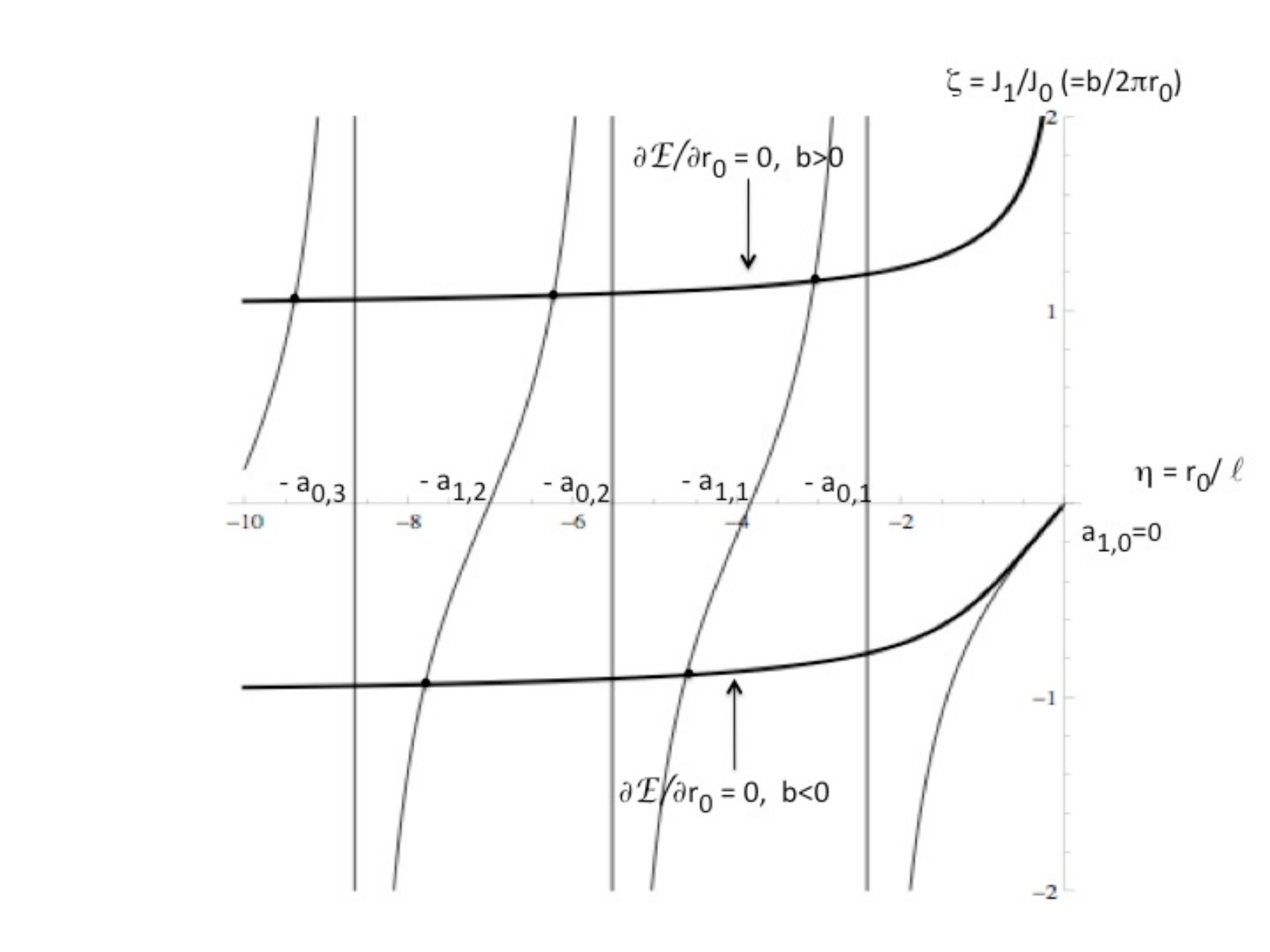} 
    \caption{\footnotesize Nonvanishing magnetic field outside the rope and continuous at the boundary, with $\ell < 0$.  Notice the change of the sign of $b$ with respect to Figure~\ref{fig3}.}
    \label{fig4}
 \end{figure} }
 
Thus, the {critical} values of $\ell_\jmath$ and $ r_{0\jmath}$ are \textit{quantized} for {$B_0$ and $b$ given. For $b$ positive,  $\eta_{\jmath} = r_{0\jmath}/\ell_\jmath$ lies in the interval $ (\mathrm{a_{1,\jmath}},\mathrm{a_{0,\jmath+1}} )$; for $b$ negative, in the interval $ (\mathrm{a_{0,\jmath}}, \mathrm{a_{1,\jmath}})$.}
The second derivative 
$\partial^2\Ecal/\partial {r_0}^2$ is positive all along the curves 
$\partial\Ecal/\partial r_0 =0 $.\footnote{
One gets, starting from Equation (\ref{e32}), that
${\partial \Ecal} / {\partial{r_0}} =  (B_0^2/4) r_0 \big(1-\zeta^2+ \zeta (\zeta+\zeta^{-1})/D \big)= 0$, where $D=1-(r_0/\ell)(\zeta+\zeta^{-1})$. Using Equation (\ref{e33}), one gets that, at a critical point $r_0$  of $\Ecal$,
\[\frac{\partial^2\Ecal}{\partial {r_0}^2} =- \frac{B_0^2}{4} \frac{2}{D \ell} \zeta^2\ (\zeta+\zeta^{-1}).\]
Using Equation (\ref{e33}), it is easy to show that $b\, D<0$, regardless of the sign of $b$. Therefore ${\partial^2\Ecal}/{\partial {r_0}^2 }>0$ at all critical points.} 
Thus the solutions are minimizers of the energy. The first interval is special. The origin is indeed a minimum, but with $r_0 = \infty$. 
  
As noted above, for large $|\eta_{\jmath}|$, we have that $r_{0,\jmath}\approx |b|/2\pi$.  Thus, for large $\jmath$, the energy Equation (\ref{e25}) is given approximately by 
  \begin{equation}
\label{e35}
\Ecal_\jmath \approx  \frac{B_0^2}{8}\frac{b^2}{4\pi^2}\left(1-\frac{2}{\eta_{\jmath}} + 2\ln \frac{2\pi R}{|b|} \right) +\frac{B_0^2}{8}R^2.
\end{equation} 
It is straightforward to obtain the following asymptotic expressions for $\eta_{\jmath}$,
\begin{equation}\label{e36}
\eta_{\jmath} \approx (\jmath-1/2)\pi - \frac{1}{2\pi(2j-1)},  \quad\jmath \gg 1, \ b > 0,\nonumber\end{equation}
\begin{equation}\label{e37}\eta_{\jmath} \approx \jmath\pi - \frac{1}{4\pi \jmath},\qquad j \gg 1, \ b < 0,
\end{equation} 
from the asymptotic expressions for Bessel functions of large argument, see \cite{abram},
\begin{equation}\label{e38}
J_\nu(x) = \sqrt{2/(\pi x)} \cos\left(x - \frac12\nu \pi - \frac14\pi\right) + O(1/|x|), \ \ |x| \gg 1.
\end{equation}
The limiting value of the energy $\Ecal_\jmath$ as $\jmath\rightarrow \infty$ is given by
 \begin{equation}
\label{e39}
\Ecal_\infty=  \frac{B_0^2}{8}\frac{b^2}{4\pi^2}\left(1 + 2\ln \frac{2\pi R}{|b|} \right) +\frac{B_0^2}{8}R^2.
\end{equation}

Tables~\ref{table1} and \ref{table2} present some numerical values attached to the flux tubes, both for $\ell>0$. The last column relates to the magnetostatic energy, except for the logarithmic term, which  varies more slowly than the others, and the $R^2$ term, where $R$ may be understood as a mean distance between neighbouring ropes. The fact that this last term is so large might explain the preferred occurrence of the model of the next section, in which this term is absent the free energy.

\begin{table}[h]
\centering
 \begin{tabular}{crccc} 
\hline
$\jmath$ & $\eta_\jmath=\frac{r_{0,\jmath}}{\ell}$ &  $\zeta_j = \frac{b}{2\pi r_{0,\jmath}}$  & $\frac{2\pi\ell_\jmath}{b}$ &$(\frac{2\pi}{b})^2\ (r_0^2-\frac{b\ell}{\pi})$ \\
\hline  
2 & 4.6124 & 0.8859 &0.2447&0.7848\\
3 &7.7995  &0.9338&0.1373&0.8721\\
4& 10.9581 &0.9533&0.0957&0.9088\\
5 &14.1087  &0.9639&0.0735&0.9292\\
10 &29.8321&0.9831&0.0341&0.9665\\ 
{20}&61.2548&0.9918&{0.0165}&0.9837\\
{30}&92.6729&0.9946&{0.0108}&0.9892\\
{50}&155.5064&0.9968&{0.0065}&0.9936\\
{100}&312.5873&0.9984&{0.0032}&0.9968\\
\end{tabular}
\vspace{5pt}
 \caption{Twisted flux rope with no surface current -- the case $\ell > 0,\ b>0$.  The term in $\ln{R/b}$ is omitted, as $R$ is an unknown parameter, and the $\ln$ term is varying more slowly than the other terms.}
 \label{table1}
\end{table} 

\begin{table}[h]
\centering
\begin{tabular}{crccc} 
\hline
$\jmath$ & $\eta_\jmath=\frac{r_{0,\jmath}}{\ell}$ &  $\zeta_j = \frac{b}{2\pi r_{0,\jmath}}$  & $\frac{2\pi\ell_\jmath}{b}$ &$(\frac{2\pi}{b})^2\ (r_0^2-\frac{b\ell}{\pi})$ \\
\hline  
1 & 3.0526 & -1.1508 &-0.2850&1.3247\\
2 &6.2320  &-1.0770&-0.1490&1.1600\\
3&9.3889&-1.0520&-0.1013&1.1064\\
4 &12.5388  &-1.0391&-0.0768&1.0797\\
5&15.6855  &-1.0314&-0.0618&1.0637\\
10 &31.4044&-1.0158&-0.0314&1.0318\\ 
{20}&62.8260&-1.0079&-0.0158&1.0159\\
{30}&94.2438&-1.0053&-0.0106&1.0106\\
{50}&157.0772&-1.0032&-0.0063&1.0064\\
{100}&314.1581&-1.0016&-0.0032&1.0032\\
\end{tabular} \vspace{5pt}  
 \caption{Twisted flux rope with no surface current -- the case $\ell>0,\ b<0$.  The term in $\ln{R/b}$ is omitted.}
\label{table2}
\end{table}
  
 The magnetic flux is $B_0$ per unit area outside the core, and $B_0b\ell/\pi r_0^2 =B_0 (1-\zeta^4)$ inside the core. There is therefore no concentration of the magnetic flux. The largest \textit{positive} value of $1-\zeta^4$, namely $0.384$, obtains for $\ell>0, \ b>0, \ \jmath =2$ (Table~\ref{table1}). For this solution, the magnetic lines of force change handedness once, which of course decreases the flux total through the rope. The largest-in-magnitude \textit{negative} value of $1-\zeta^4$, namely  $-0.754$,   obtains for $\ell>0, \ b<0, \ \jmath =2$ (Table~\ref{table2}). This negative density of flux  inside the rope, which is of opposite sign to the outside flux, exceeds in modulus that of  the preceding case $b >0$. 
 In any case, the flux concentration is very small, and even absent for $\jmath > 2$.

\section{The Twisted Rope with No Field Outside}\label{trnf}

{We consider an isolated  cylindrical magnetic rope of radius $r_0$  with force-free {\color{black} axisymmetric internal field and vanishing external field
(one could also consider nonaxisymmetric fields \cite{taylor86}).}
The internal field is of the form
\begin{equation}\label{e40}
B_\theta(r) = A J_1(r/\ell), \quad B_z(r) = A J_0(r/\ell), 
\end{equation} 
where $\ell$ is the (constant) Beltrami parameter.  It is convenient to express the parameters $A$ and $\ell$ in terms of physical parameters, namely $B_0$, the flux per unit area just inside the boundary, and $\Ical$, the current inside the rope.  The relations are then the same as those obtained in the preceding sections, namely\begin{equation}\label{e41}
  A = B_0 J_0(r_0/\ell), \quad
\frac{J_1}{J_0}(r_0/\ell) = \frac{b}{2\pi r_0},
\end{equation}
where
\begin{equation}\label{e42}
b = \frac{4\pi \Ical}{B_0 c}.
\end{equation} 
The discontinuity in the field across the boundary requires a surface current density, which is given by
\begin{equation}\label{e43}
 j_\theta = -\frac{c}{4\pi} B_0, \quad
j_z  = - \frac{\Ical}{2\pi r_0}, \end{equation} 
and yields a jump in the isotropic magnetic pressure, which is inside the boundary equal to $|\Bvec(r_0)|^2/8\pi$, and vanishes outside. 
This is balanced by the plasma pressure.      
{\color{black}We do not consider this problem here, and content ourselves with the topological properties, which are fully apparent.}

This model is commonplace in standard treatments of magnetic ropes. The new feature here, as in the preceding section, is an additional constraint which derives from an underlying helicoidal topology.  In addition to the interior current [$\Ical$], we regard the pitch of the field-line helices as an invariant.  Note that, as the external field vanishes, the helicoidal topology is virtual, and should be understood as arising from the limit of field configurations in which the ratio of the external to internal field tends to zero  at constant pitch (so that the helical character is preserved). Alternatively, the helical structure may be derived from the vector potential, as follows.  Inside the core, the vector potential {\color{black}is $\ell \Bvec$ (\textit{cf.} Equation (\ref{e16})), which is divergenceless}.  Therefore, by continuity, the vector potential outside the core is given by $\Avec = \ell (0,B_0 b/(2\pi r),B_0)$.}\footnote{The vector potential being continuous amounts to saying that
there are no $\delta$-function singularities in the $\Bvec$-field.
Consider \textit{e.g.} the {\it discontinuous} vector potential
$$\Avec(r,\theta,z) = 0,         \quad             r < r_0,
\qquad \Avec(r,\theta,z)= C \grad \theta, \quad r > r_0,$$
then $\Bvec = \rot \Avec =C/(2\pi r_0) \delta(r- r_0) \bf{\hat z} .$  That is, one gets a $\delta$-function $\Bvec$-field on the cylinder $r = r_0.$} 
{As the  external vector potential is irrotational, it may be expressed, locally at least, as a gradient  [$\grad \phi$], with $\phi = B_0\ell (z+(b/2\pi)\theta)$.  The level sets of $\phi$ are then  helicoids of pitch $b$. In this way, the twisted rope may be viewed as the core (in the sense of defect theory) of a helicoidal topology with vanishing external field.  (Note that under a change of gauge $\Avec \rightarrow \Avec + \grad f$, the scalar field $\phi$ becomes $\phi + f$.  The helicoidal level sets are deformed, but as $f$ is necessarily regular ($\Avec$ is required to be continuous), their topological structure, and in particular the core, persists.) Taking $B_0$ and $b$ to be invariants of the magnetic rope is then equivalent to taking the current and pitch as invariants.

The energy in this case is calculated similarly to Equation (\ref{e25}), and is given by
 \begin{equation}
\label{e44}
\Ecal
=\frac{B_0^2  b^2}{16\pi^2} \left( 1 +\zeta^{-2} -\frac{2\pi\ell}{b} \right), \quad \mathrm{where }\ \zeta = \frac{b}{2\pi r_0}.\end{equation}

The extrema 
are characterised by the condition $\partial \Ecal/\partial r_0 = 0$.   With $\partial \ell/\partial r_0$ given by  Equation (\ref{e31}),
this may be written as

  \begin{equation}
\label{e45}
\zeta^4 +\left(3-\frac{b}{\pi\ell}\right) \zeta^2 -\frac{b}{\pi\ell}=0,
\end{equation}
or  equivalently,
 \begin{equation}
\label{e46}
 \eta=\frac{1}{2} \zeta +\frac{\zeta}{1+\zeta^2},\quad\mathrm{where }\ \eta = r_0/\ell.
\end{equation}
From Equation (\ref{e41}), we have  the consistency condition
\begin{equation}\label{e47}
\zeta = \frac{J_1}{J_0}(\eta).
\end{equation}
Critical points are then given by simultaneous solutions of Equation (\ref{e46}) and Equation (\ref{e41}).
These are shown in Figure~\ref{fig5}, and numerical values are presented in Table~\ref{table3}.
In each interval $\mathrm a _{1,\jmath-1} < r_0/\ell < \mathrm a _{0,\jmath}$, there is a single critical point [$r_{0,\jmath}$] with corresponding Beltrami parameter {\color{black} $\ell_\jmath$.}

In fact, all critical points of $\Ecal$ are local minima.  A straightforward calculation yields, after some algebra,
the following expression for the second derivative of the energy evaluated at the critical points:
  \begin{equation}
\label{e48}
\frac{\partial ^2 \Ecal }{\partial{r_0}^2} = \frac{B_0^2}{2} \frac{\zeta ^2 - 3}{\zeta^2 + 1}.
\end{equation}
It is readily verified from Table~\ref{table3} that $\zeta^2 > 3$ for all critical points.

For large $\jmath$, $\eta_j$ is large and $\zeta_j \approx 2\eta_j$.  From the asymptotic expression for $a_{0\jmath}$ (which follows from Equation (\ref{e38}) and the fact that $J_0' = -J_1$), one readily obtains
\begin{equation}
\label{e49}
r_{0,\jmath} \approx \frac{b}{4\pi^2 \jmath}, \quad \ell_\jmath\approx \frac{b}{4\pi^3 \jmath^2}. 
\end{equation}
Thus, for large $\jmath$, the radius of the flux rope goes to zero even as the number of oscillations in the handedness of the field increases.  
 \begin{figure}[h] 
   \centering
   \includegraphics[width=4.in]{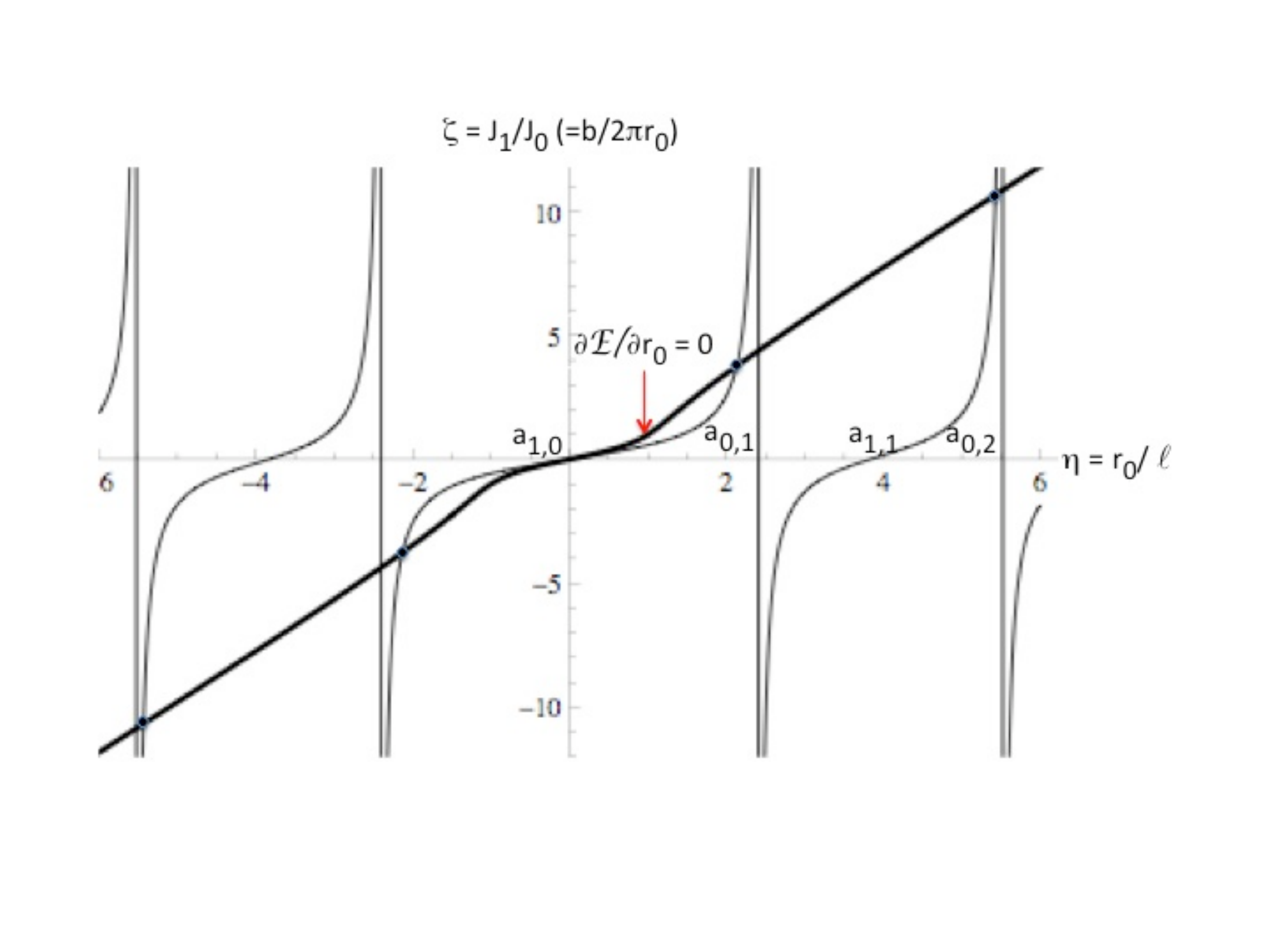} 
   \caption{\footnotesize No magnetic field outside the rope; the two first solutions for $\ell>0$ and $\ell<0$, corresponding respectively to $b>0$ (right-handed magnetic-field lines of force) and $b<0$ (left-handed lines of force). See Table~\ref{table3} for numerical values.}
   \label{fig5}
\end{figure}
The energy Equation (\ref{e44}) in this limit  is given by
\begin{equation}
\label{e50}
\Ecal_j \approx \frac{B_0^2b^2}{16\pi^2}\left(1-\frac{1}{4\mathrm a _{0,\jmath}^2)}\right)\approx \frac{B_0^2b^2}{16\pi^2}\left(1-\frac{1}{4\pi^2\jmath^2}\right).
\end{equation}
\begin{table}[h]
\centering
\begin{tabular}{ccccc} 
\hline 
$\jmath$ & $\eta_\jmath=\frac{r_{0,\jmath}}{\ell}$ &  $\zeta_j = \frac{b}{2\pi r_{0,\jmath}}$  & $\frac{2\pi\ell_\jmath}{b}$ & $1+(\frac{2\pi r_{0,\jmath}}{b})^2-\frac{2\pi \ell_\jmath}{b}$ \\
\hline  
1 & 2.1288 & 3.7610 &0.1249&0.9458\\
2 &5.4258  &10.6657&0.0173&0.9915\\
3& 8.5950 &17.0733&0.0068&0.9966\\
4 &11.7488  &23.4123&0.0036&0.9981\\
5&14.8973  &29.7273&0.0023&0.9989\\
{10}&30.6183&61.2039&0.00053&0.99973\\
{20}&62.0404&124.0647&0.00013&0.99994\\
{30}&93.4584&186.9060&5.725$\times10^{-5}$&0.99997\\
{50}&156.2918&312.5773&2.047$\times10^{-5}$&0.999989\\
{100}&313.3727&626.7421&4.0916$\times10^{-6}$&0.999997\\ 
\end{tabular}\vspace{5pt} \caption{Twisted flux rope with no external field.  The last column is the line energy, to a factor $\frac{1}{16\pi^2}B_0^2  {b^2}$.} 
\label{table3} \end{table}
The energy scales as $b^2$, as expected, and increases but slightly when $\jmath$ increases, and thus does not depend strongly on the radius of the rope. The energy of the flux rope with vanishing external field is clearly smaller than the case considered in the preceding section, where the surface current was taken to vanish and $\Bvec$ approached the uniform field $B_0\,{\bf \hat z}$, even if the energy of the uniform field $B_0\,{\bf \hat z}$ is neglected.  

  The magnetic-flux density in the core is $B_0b\ell/\pi r_0^2 = 4B_0 (1+\zeta^2)/(3+\zeta^2)$. Thus, for a given amplitude $B_0$, the flux density tends very quickly to $4B_0$ for $\zeta\rightarrow \infty$; in fact it is already $\approx 3.53B_0$ for the first quantized state $\jmath = 1$. In this state the magnetic lines of force do not change handedness in the core. But the flux concentration increases even when $\jmath$ increases, \textit{i.e.} for states whose handedness changes sign $2( \jmath  -1)$ times. }
  
In the preceding calculation, the twisted rope has been treated as the limit of a rope with $B_0 \rightarrow 0$ with fixed interior current. In a forthcoming publication this assumption will be discussed  and compared to an alternative, namely  a flux rope with fixed internal magnetic flux.
  
  \section{Discussion}\label{s8}
The foregoing gives a very idealised picture of a flux tube, even if it {\color{black}correctly} captures the singular nature of a tube of current, \textit{i.e.}~that the current is a singularity of the magnetic field, {\color{black} or, in the case of a twisted rope with no field outside, an irrotational vector potential}. In the applications, we have restricted ourselves to cases where the physical invariant is the current intensity [$B_0b$], which stems from the no-divergence property of the current density {in the static case}. However, the results obtained are of some importance in the analysis of the physics of flux ropes, in particular those where the magnetic field is confined ({\color{black}no magnetic field outside the core,} Section~\ref{trnf}), to which we restrict attention in what follows.\footnote{Remember that a magnetic vortex line in a Type II superconductor is also the singularity of an irrotational vector potential; the magnetic field is expelled from the specimen, except in the core region that carries a quantized magnetic flux \cite{pgg89}.} {\color{black}A number of solar observations of coronal loops seem to indicate that segments of constant {cross-sectional radius}, implying constant $b$, can exist along long distances; see figures in \cite{aschwanden00}.} Also, according to \cite{bellan03}, flux ropes relax after helicity injection to a state of axially invariant cross-section. Our model is consistent with these observations.

{\color{black} {The} equations that we obtain are based on a minimization process at constant current 
(\textit{cf.}~Equation (\ref{e5})) }
\begin{center}
$\Ical =(c/4\pi)B_0b$.
\end{center} 
But a flux rope also carries two other invariants, namely the magnetic flux 
(\textit{cf.}~Equation (\ref{e20})) 
\begin{center}
$\Phi=B_0b\ell$,
\end{center} and the helicity $\Hcal$ per unit length. This latter quantity can be written \begin{equation}
\label{e51}
\Hcal= \int \Avec\cdot\Bvec r \mathrm{d} r \,\mathrm{d}\theta = 8\pi\ell\Ecal,
\end{equation} 
{as we may take  $\Avec = \ell \Bvec$ for a force-free field.} These three invariants are not linearly related, whereas they add linearly when several ropes gather to {\color{black} form a single}  rope. Therefore any phenomenon implying a splitting or a merging of ropes is necessarily attended by dynamical processes, which result in the modification of some or all the invariants. We briefly discuss a few examples, where two of the three invariants are conserved {\color{black} during such dynamical processes as reconnection or splitting.}\vspace{5pt}

 i) \textit{$\Ical,\ \Hcal$ conserved.} We assume that a $\jmath$-rope, with $\jmath$ large, current intensity {\color{black} $(c/4\pi)B_{(\jmath)} b_\jmath$}, is split into $\jmath$ one-ropes belonging to the $\jmath = 1$ state, each with current intensity $(c/4\pi)B_{(1)} b_1$. The conservation conditions can be written:
\begin{equation}
\label{e52}
\jmath B_{(1)}b_1=B_{(\jmath)} b_\jmath, \quad \ell_\jmath (B_{(\jmath)} b_\jmath)^2 =\alpha \jmath \ell_1 (B_{(1)} b_1)^2,\ \alpha \approx 0.9458
\end{equation}  
  Notice that, according to Table~\ref{table3}, the energy per unit {\color{black} length} is not very different from one state to the other, for $\jmath$ large enough; hence the coefficient $\alpha$ {is nearly unity}.
Comparison of the two conservation laws yields \begin{center}
$ \ell_1 =(\jmath /\alpha)\ell_\jmath$.
\end{center} One finds, {\color{black} using~Equation (\ref{e44}) for $\eta_j$ and Table~\ref{table3} for $\eta_1$,} that\begin{center}
$r_1 \approx 0.7164\ r_\jmath$
\end{center} so that the {\color{black} radius of the subdividing ropes is not much different from the  radius of the original} rope. Finally, assuming $B_{(\jmath)} = B_{(1)}$, the Burgers' vector of the subdividing ropes is $\jmath$ times smaller than the Burgers' vector of the {\color{black} original} rope. 
   
   The energies are considerably reduced in this process; the  energy $\Ecal$ after splitting is:\begin{center}
 $\Ecal=\jmath\ \Ecal_1 = 
({\alpha}/{\jmath})\ \Ecal_\jmath$,
\end{center} which favors the process, but at the expense of a large increase of the flux; one finds \begin{center}
$\Phi = \jmath\ \Phi_1 =({\jmath}/{\alpha})\ \Phi_\jmath$.\\ 
\end{center}

 ii)
  \textit{$\Phi,\ \Ical$ conserved. }The conservation conditions can be written:
\begin{equation}
\label{e53}
\jmath \ell_1 B_{(1)}b_1=\ell_\jmath B_{(\jmath)} b_\jmath, \quad \jmath  B_{(1)}b_1=B_{(\jmath)} b_\jmath.
\end{equation}  These equations yield \begin{center}
$\ell_1= \ell_\jmath$
\end{center} and a decrease of energy \begin{center}
$\Ecal = \jmath\ \Ecal_1 =(\alpha/\jmath)\ \Ecal_\jmath$.
\end{center} This process is thus feasible, at the expense of a change of helicity \begin{center}$\Hcal = \jmath\ \Hcal_1= (\alpha/\jmath)\ \Hcal_\jmath$.\end{center} Notice that, using~Equation (\ref{e37}) for $\eta_j$ and Table~\ref{table3} for $\eta_1$, one finds 
\begin{center}$r_1 \approx (0.6965/\jmath)\ r_\jmath$.\\\end{center} 

iii) \textit{$\Hcal,\ \Phi$ conserved.} This is the only case out of the three considered here where the splitting is not favored; on the contrary, one expects that the elementary ropes merge to form a larger one, with a large dissipation of energy. This possibility might be {\color{black} related to a question raised by Parker \cite{parker09}, namely ``what causes  the subsurface convection \textrm{of the Sun}  to sweep thousands of magnetic fibrils together and then to compress them into two or more kilogauss?''  }Note further that the conservation of helicity is an assumption often made in high electrical conductivity media \cite{berger84}, {\color{black} while the conservation of flux is also physically reasonable}. 

Start from a $\jmath$-rope split into $\jmath$ one-ropes. The conservation conditions can be written:
\begin{equation}
\label{e54}
\jmath \ell_1 B_{(1)}b_1=\ell_\jmath B_{(\jmath)} b_\jmath, \quad \ell_\jmath (B_{(\jmath)} b_\jmath)^2 =\alpha \jmath \ell_1 (B_{(1)} b_1)^2,\ \alpha \approx 0.9458
\end{equation}  
  
   The comparison of the two conservation laws yields \begin{center}
$ \ell_1 =(\alpha/\jmath)\ell_\jmath$.
\end{center} This is the inverse of the relation in the previous case, and one finds indeed that the energy is considerably increased in this process
   \begin{center}$\Ecal = \jmath \Ecal_1 = 
(\jmath/\alpha) {\Ecal_\jmath}$,\end{center} which is thus not favored.
On the other hand, the gathering of one-ropes into large $\jmath$-ropes is {\color{black} energetically favoured}, in a process in which the current intensity varies, which means the appearance of transient displacement currents, \textit{i.e.} transient electric fields. Using~Equation (\ref{e37}) for $\eta_j$ and Table~\ref{table3} for $\eta_1$, one obtains \begin{center}
$r_\jmath \approx 1.5603 \ \jmath^2 r_1$. \end{center} \smallskip

These are of course ideal processes, which demand to be explored theoretically for their own sake. A real process would certainly be a mixture of these, not to mention the well-known kink processes that {\color{black} produce writhing in the flux ropes}. 

{\color{black} A number of numeric simulations of {\color{black} single-fluid MHD} have been performed, all using models of twisted flux ropes with force-free fields and conservation of helicity, (\textit{e.g.} \cite{linton06,torok11}), in order to better understand the phenomena of reconnection (not splitting) which yield flares, prominences, \textit{etc}, in the Sun's corona in relation with the huge energy which can be released in these processes. Various modes of reconnection have been recognized, depending on the relative signs of the twist, the relative directions of the mean field [$\bvec$], \textit{etc}. In this discussion we have limited ourselves to parallel tubes of the same twist. It would be interesting to extend our calculations to other cases. It would also be interesting to compare the quantitative features of the flux ropes obtained from the foregoing analysis -- for example, the structure of the interior and surface currents, and their relation to the radius of the flux tube -- with the structures that emerge from numerical simulations in the high-field regime.}
 \bigskip
   
 \section*{Appendix: Some Fundamentals of Defect Theory}\label{deftheo} The key ingredient of the theory of line defects in an {\it ordered medium} $\Omega$ (a crystal) is the \textit{Volterra process}, see \cite{friedeldisloc}; it describes the creation of a line singularity characterized by an isometry of the medium, namely a translational symmetry $\bvec$ and/or a rotational symmetry $\bm\omega=\omega \bm\nu$, $\omega$ being the angle of rotation, $\bm\nu$ its axis.

\begin{figure}[h] 
   \centering
   \includegraphics[width=4.in]{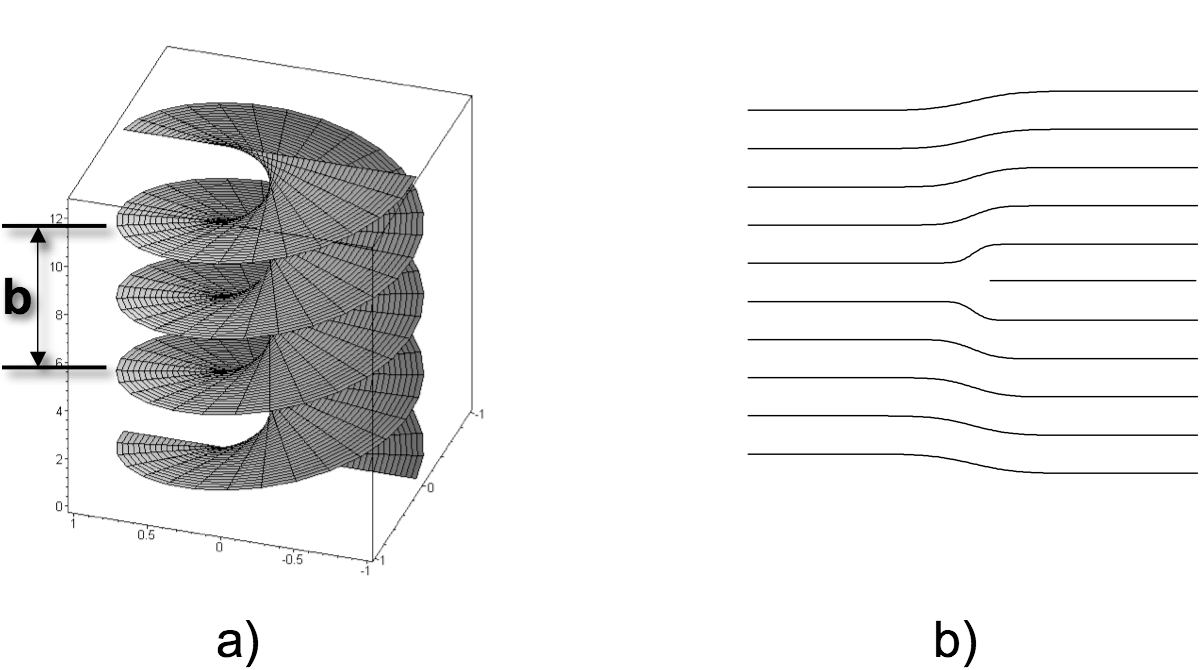} 
   \caption{\footnotesize Elementary dislocations (displacement $\dvec (\rvec )$ reduced to
a pure translation) in a smectic phase: a) screw dislocation, $\bvec$ is in the direction of $L$; b) edge dislocation, $\bvec$ is perpendicular to $L$. In both cases $L$ is an infinite straight line.}
   \label{fig6}
\end{figure}

 Consider in a perfectly ordered $\Omega$ a line $L$, and cut the crystal along an element of surface $\Sigma$ bound by $L$ ($L = \partial \Sigma$ is a loop or an infinite line); $\Sigma_1$ and $\Sigma_2$ are the two lips of $\Sigma$, the \textit{cut surface}.  
Displace the two lips $\Sigma_1$ and $\Sigma_2$ by a relative rigid displacement $\bvec$ and a rotation $\bm\omega $, \textit{viz.} by a total displacement (we assume $\omega$ small):
 \begin{equation}
\label{e55}
\dvec (\rvec ) = \bvec+\bm\omega \times \rvec,
\end{equation} 
$\bvec$ is \textit{the Burgers' vector} of the line defect, $\bm\omega$ its \textit{rotation vector} \cite{klfr08}.
  Introduce now a perfect wedge of matter in the void left by the displacement, in order to fill it with exact contacts with the lips (which is possible since $\dvec (\rvec )$ is an isometry) $-$ or remove matter in the regions of double covering. Then reestablish the bonds at the lips $\Sigma_1$ and $\Sigma_2$ and let the medium relax elastically.  The line defect thus obtained is called a \textit{dislocation} if $\omega = 0$, a \textit{disclination} if $\bvec = 0$. Figure~\ref{fig6} shows dislocations in a 
smectic phase, in which there is only one finite repeat distance $d$ perpendicular to the layers (here $|\bvec|=2d$). A \textit{screw dislocation} is when $\bvec$ is along the line $L$ (here along the axis of the helically transformed layers); an \textit{edge dislocation} when $\bvec$ is orthogonal to $L$ $-$ here $\bvec$ is in the plane of the drawing, $L$ perpendicular to it, $|\bvec|=d$.

Such a process has of course no meaning along $L$ itself, and one assumes henceforth that a toric region has been removed along this line. This ``core" is a region where the above process is not valid. The Volterra process undoubtedly
introduces internal stresses in an elastic medium, but these stresses are non-singular on $\Sigma_1$ and $\Sigma_2$ insofar as $\dvec (\rvec )$ obeys Hookean elasticity. Henceforth, the final result does not depend on the exact choice of $\Sigma$ .\\

\textit{Vortex lines}: this term refers to a line defect $L$ whose characteristic invariant is not an element of symmetry of the medium as above but the \textit{circulation} $\gamma =\oint_C \uvec.\mathrm{d}\svec$ of a vector $\uvec$. In a irrotational fluid this would be the flow velocity; in a superconductor the vector potential. Notice that the vector $\uvec$ itself has the geometry of a disclination, since it rotates by $2\pi$ about the line $L$. One will have recognized that the singularities investigated in the present work are of the vortex line type, but display a geometry locally reminiscent of a screw dislocation in a smectic.

 \section*{Acknowledgements}{\footnotesize We are grateful to Jean-Louis Le Mou\"el for a careful reading of the manuscript and useful remarks. JMR thanks the Institut de Physique du Globe for hospitality when this work was started.
  This is IPGP contribution \# 3382.}
  

     \bibliographystyle{acm}


\end{document}